# Isomorphism between the multi-state Hamiltonian and the second-quantized many-electron Hamiltonian with only 1-electron interactions

Jian Liu [1, a)]

1. *Beijing National Laboratory for Molecular Sciences, Institute of Theoretical and Computational Chemistry, College of Chemistry and Molecular Engineering, Peking University, Beijing 100871, China*

a) Electronic mail: jianliupku@pku.edu.cn

We introduce the isomorphism between the multi-state Hamiltonian and the second-quantized many-electron Hamiltonian (with only 1-electron interactions). This suggests that all methods developed for the former can be employed for the latter, and vice versa. The resonant level (Landauer) model for nonequilibrium quantum transport is used as a proof-of-concept example. Such as the classical mapping models for the multi-state Hamiltonian proposed in our previous work [J. Chem. Phys. (submitted)] lead to exact results for this model problem. We further demonstrate how these methods can also be applied to the second-quantized many-electron Hamiltonian even when 2-electron interactions are included.

## I. Introduction

Consider a Hamiltonian operator for $F$ orthonormal states

$$\hat{H} = \sum_{m,n=1}^{F} H_{mn} |m\rangle\langle n| \quad . \tag{1}$$

The Hamiltonian matrix is often a real symmetric one, where $H_{nm} = H_{mn}$. By extending Schwinger's oscillator model of angular momentum[1, 2], state $|n\rangle$ can be mapped[3] as

$$|n\rangle \mapsto \underbrace{|0_1, 0_2, \cdots, 1_n, \cdots, 0_F\rangle}_{F-\text{states}} \tag{2}$$

such that it is viewed as a single excitation from the vacuum state $|\bar{0}\rangle \mapsto \underbrace{|0_1, 0_2, \cdots, 0_n, \cdots, 0_F\rangle}_{F-\text{states}}$, i.e.,

$$|n\rangle = \hat{a}_n^+ |\bar{0}\rangle \quad . \tag{3}$$

Here an excitation represents the occupation of the corresponding state. The vacuum state $|\bar{0}\rangle$ is orthogonal to any occupied state $|n\rangle$. As suggested in Paper I[4], the creation and annihilation operators for Eq. (3) can be defined as

$$\begin{aligned} \hat{a}_n^+ &= \underbrace{|0_1, 0_2, \cdots, 1_n, \cdots, 0_F\rangle}_{F-\text{states}} \underbrace{\langle 0_1, 0_2, \cdots, 0_n, \cdots, 0_F|}_{F-\text{states}} = |n\rangle\langle\bar{0}| \\ \hat{a}_n &= \underbrace{|0_1, 0_2, \cdots, 0_n, \cdots, 0_F\rangle}_{F-\text{states}} \underbrace{\langle 0_1, 0_2, \cdots, 1_n, \cdots, 0_F|}_{F-\text{states}} = |\bar{0}\rangle\langle n| \end{aligned} \quad . \tag{4}$$

The combination of a pair of creation and annihilation operators are

$$\begin{aligned} &\hat{a}_n^+\hat{a}_n = |n\rangle\langle n|, \ \hat{a}_m^+\hat{a}_n = |m\rangle\langle n|, \\ &\hat{a}_n\hat{a}_n^+ = \hat{a}_m\hat{a}_m^+ = |\bar{0}\rangle\langle\bar{0}| \\ &\hat{a}_n\hat{a}_m^+ = \hat{a}_m\hat{a}_n^+ = 0 \quad (n \neq m) \end{aligned} \quad . \tag{5}$$

Substituting Eq. (5) into Eq. (1) leads to

$$\hat{H} = \sum_{m,n=1}^{F} H_{mn} \hat{a}_m^+ \hat{a}_n \quad . \tag{6}$$

Here each state is described by a degree of freedom. It is straightforward to show the anticommutation relations

$$\left[\hat{a}_n^+, \hat{a}_m^+\right]_+ = 0, \quad \left[\hat{a}_n, \hat{a}_m\right]_+ = 0,$$
$$\left[\hat{a}_n^+, \hat{a}_m\right]_+ = \begin{cases} |n\rangle\langle n| + |\bar{0}\rangle\langle\bar{0}| & (n = m) \\ |n\rangle\langle m| & (n \neq m) \end{cases} \quad . \tag{7}$$

Similarly, commutation relations are also naturally derived[4]. These relations suggest that the underlying degrees of freedom are neither bosons nor fermions[4]. A novel theoretical framework is then developed for constructing equivalent expressions of the Hamiltonian operator in quantum mechanics and their mapping models in the Cartesian phase space[4]. It also offers a new way[4] to derive the seminal Meyer-Miller semiclassical mapping model other than the original ones proposed by Meyer and Miller[5] and by Stock and Thoss[3].

The purpose of the paper is to introduce an isomorphism between the multi-state Hamiltonian [Eq. (1)] and the second-quantized many-electron Hamiltonian (with only 1-electron interactions). This suggests that all methods developed for the former can naturally be employed for the latter, and vice versa. The paper is organized as follow. Section II first expresses the creation and annihilation operators in terms of the occupation number (ON) vectors for the second-quantized many-electron Hamiltonian. The isomorphism is then presented when only 1-electron interactions exist in the second-quantized many-electron Hamiltonian. Section III briefly reviews several classical mapping models for the multi-state Hamiltonian [Eq. (1)], which are employed to study the nonequilibrium dynamics of the second-quantized many-electron Hamiltonian in Section IV. The resonant level (Landauer) model[6, 7] and the model system consisted of two electrons and four spin-orbitals in Ref. [8] are used as two principle-of-concept examples. Section V summarizes and concludes.

## II. Isomorphism between the multi-state Hamiltonian and the second-quantized many-electron Hamiltonian with only 1-electron interactions

Consider the general second-quantized many-electron Hamiltonian

$$\hat{H}_{el} = \sum_{m,n=1}^{F} h_{mn}\hat{\alpha}_m^+\hat{\alpha}_n + \frac{1}{2}\sum_{l,m,n,j=1}^{F} g_{ljmn}\hat{\alpha}_l^+\hat{\alpha}_m^+\hat{\alpha}_n\hat{\alpha}_j \quad , \tag{8}$$

where $F$ is the total number of orthogonal spin-orbitals, $h_{mn}$ and $g_{ljmn}$ are 1- and 2-electron integrals, respectively, and $\{\hat{\alpha}_n^+, \hat{\alpha}_n\}$ are the creation and annihilation operators for the individual spin-orbitals[9, 10]. An ON vector $|\mathbf{k}\rangle$ is

$$|\mathbf{k}\rangle = |k_1, k_2, \ldots, k_F\rangle, \ k_j = \begin{cases} 1 & j\text{-th spin-orbital occupied} \\ 0 & j\text{-th spin-orbital unoccupied} \end{cases} \quad . \tag{9}$$

$\{\hat{\alpha}_n^+\}$ are defined by the relations

$$\begin{aligned} &\hat{\alpha}_n^+ |k_1, k_2, \ldots, 0_n, \ldots, k_F\rangle = \Gamma_n^{\mathbf{k}} |k_1, k_2, \ldots, 1_n, \ldots, k_F\rangle \\ &\hat{\alpha}_n^+ |k_1, k_2, \ldots, 1_n, \ldots, k_F\rangle = 0 \end{aligned} \tag{10}$$

with the phase factor

$$\Gamma_n^{\mathbf{k}} = \prod_{j=1}^{n-1} (-1)^{k_j} \tag{11}$$

that is equal to +1 if there are an even number of electrons in the spin-orbitals $j < n$ (i.e., preceding $n$ in the ON vector) and equal to -1 if there are an odd number of electrons in these spin-orbitals[9, 10]. It is heuristic to express the creation operator in terms of the ON vectors

$$\hat{\alpha}_n^+ = \sum_{k_1=0}^{1} \cdots \sum_{k_{n-1}=0}^{1} \sum_{k_{n+1}=0}^{1} \cdots \sum_{k_F=0}^{1} \Gamma_n^{\mathbf{k}} |k_1, k_2, \ldots, 1_n, \ldots, k_F\rangle\langle k_1, k_2, \ldots, 0_n, \ldots, k_F| \quad . \tag{12}$$

Similarly, the corresponding annihilation operator is

$$\hat{\alpha}_n = \sum_{k_1=0}^{1} \cdots \sum_{k_{n-1}=0}^{1} \sum_{k_{n+1}=0}^{1} \cdots \sum_{k_F=0}^{1} \Gamma_n^{\mathbf{k}} |k_1, k_2, \ldots, 0_n, \ldots, k_F\rangle\langle k_1, k_2, \ldots, 1_n, \ldots, k_F| \quad . \tag{13}$$

It is easy to derive the anticommutation relations

$$\left[\hat{\alpha}_n^+,\hat{\alpha}_m^+\right]_+ = 0,\ \left[\hat{\alpha}_n,\hat{\alpha}_m\right]_+ = 0,\ \left[\hat{\alpha}_n^+,\hat{\alpha}_m\right]_+ = \delta_{nm} \quad . \tag{14}$$

It is important to note that the proof of the last equation of Eq. (14) (i.e., $\left[\hat{\alpha}_n^+,\hat{\alpha}_m\right]_+ = 0$ for $n \neq m$) requires the existence of $\left|k_1,\ldots,k_{n-1},1_n,k_{n+1},\ldots,k_{m-1},1_m,k_{m+1},\ldots,k_F\right\rangle$.

The ON vectors in Eqs. (9), (10), (12), (13) constitute an orthonormal basis in the $2^F-$ dimensional Fock space $Fk(F)$, which may be expressed as a sum of subspaces $Fk(F,N)$, i.e.,

$$Fk(F) = Fk(F,0) \oplus Fk(F,1) \oplus \cdots \oplus Fk(F,F) \quad . \tag{15}$$

Here $Fk(F,N)$ involves all ON vectors for which the sum of the occupation numbers in Eq. (9) is $N$:

$$N = \sum_{j=1}^{F} k_j \quad . \tag{16}$$

I.e., the ON vectors of the Fock subspace $Fk(F,N)$ are obtained by distributing $N$ electrons among the $F$ spin-orbitals[9, 10], the total number of which is equal to the binomial coefficient $\begin{pmatrix} F \\ N \end{pmatrix}$.

When all 2-electron integrals $g_{ljmn}$ are zero, Eq. (8) is reduced to

$$\hat{H}_{el} = \sum_{m,n=1}^{F} h_{mn}\hat{\alpha}_m^+\hat{\alpha}_n \quad . \tag{17}$$

Consider that the system can be fully described by 1-electron states—only 1-electron interactions are involved. That is, the relevant ON vectors are the ones with only 1 electron $\left|0_1,0_2,\ldots,1_n,\ldots,0_F\right\rangle$ $(n = 1,\cdots,F)$ and the true vacuum state $\left|0_1,0_2,\ldots,0_n,\ldots,0_F\right\rangle$. The related space is $Fk(F,1) \oplus Fk(F,0)$. Eqs. (12)-(13) then become

$$\begin{aligned}\hat{\alpha}_n^+ &= |0_1, 0_2, \ldots, 1_n, \ldots, 0_F\rangle\langle 0_1, 0_2, \ldots, 0_n, \ldots, 0_F| \\ \hat{\alpha}_n &= |0_1, 0_2, \ldots, 0_n, \ldots, 0_F\rangle\langle 0_1, 0_2, \ldots, 1_n, \ldots, 0_F|\end{aligned} \quad . \tag{18}$$

It is easy to verify that the anticommutation relations are

$$\begin{aligned}&\left[\hat{\alpha}_n^+, \hat{\alpha}_m^+\right]_+ = 0, \quad \left[\hat{\alpha}_n, \hat{\alpha}_m\right]_+ = 0, \\ &\left[\hat{\alpha}_n^+, \hat{\alpha}_m\right]_+ = \begin{cases} \hat{\mathbf{1}}^{(n)} & (n = m) \\ \hat{\alpha}_n^+ \hat{\alpha}_m & (n \neq m) \end{cases}\end{aligned} \quad , \tag{19}$$

with the identity operator

$$\begin{aligned}\hat{\mathbf{1}}^{(n)} &= |0_1, 0_2, \ldots, 1_n, \ldots, 0_F\rangle\langle 0_1, 0_2, \ldots, 1_n, \ldots, 0_F| \\ &\quad + |0_1, 0_2, \ldots, 0_n, \ldots, 0_F\rangle\langle 0_1, 0_2, \ldots, 0_n, \ldots, 0_F|\end{aligned} \tag{20}$$

and

$$\hat{\alpha}_n^+ \hat{\alpha}_m = |0_1, 0_2, \ldots, 1_n, \ldots, 0_m, \ldots, 0_F\rangle\langle 0_1, 0_2, \ldots, 0_n, \ldots, 1_m, \ldots, 0_F| \quad . \tag{21}$$

The subtle difference between Eq. (14) and Eq. (19) indicates that the underlying degrees of freedom of Eqs. (18)-(19) do *not* obey Fermi statistics.

Comparison of Eq. (6) to Eq. (17), that of Eq. (4) to Eq. (18), and that of Eq. (7) to Eq. (19) suggest isomorphism between the multi-state Hamiltonian [i.e., Eq. (1)] and the second-quantized many-electron Hamiltonian (with only 1-electron interactions) [i.e., Eq. (17)]. This indicates that all methods developed for the former can be employed for the latter, and vice versa.

When 2-electron interactions are included in the second-quantized many-electron Hamiltonian, the anti-commulation relations are Eq. (14) instead of Eq. (19)—the isomorphism then no longer exists. One may express the second-quantized many-electron Hamiltonian [Eq. (8)] by using many-electron states such that a multi-state Hamiltonian [Eq. (1)] is constructed for the system. All methods developed for the multi-state Hamiltonian are then still useful to study such as nonequilibrium quantum dynamics of the second-quantized many-electron Hamiltonian [Eq. (8)].

## III. Classical mapping models

Paper I introduces six different classical mapping models for the multi-state Hamiltonian [Eq. (1)]. We briefly review Models I-IV of Paper I[4] and how the initial condition of the classical trajectory is constructed for each classical mapping model. (Models V-VI of Paper I[4] can be applied in a similar way, which are not discussed in the present paper.)

### 1. Mapping model I

Employing the analogy with the classical angular momentum, one obtains the classical Hamiltonian of Model I for the multi-state Hamiltonian[4]

$$H = \sum_{n,m=1}^{F} H_{nm} \left( x^{(n)} p_y^{(m)} - y^{(n)} p_x^{(m)} \right) \quad . \tag{22}$$

Model I is reminiscent of the semiclassical second-quantized many-electron Hamiltonian proposed by Li and Miller[8]. It is not a surprise. The isomorphism introduced in Section II suggests that any methods developed for the multi-state Hamiltonian [i.e., Eq. (1)] can be applied to the second-quantized many-electron Hamiltonian (with only 1-electron interactions) [i.e., Eq. (17)], and vice versa. When Li and Miller proposed their mapping model for the second-quantized many-electron Hamiltonian, they stated that it "incorporates the anti-commutation properties of the fermionic operators" (i.e., Eq. (14))[8]. The isomorphism, however, illustrates that the mapping model has the anti-commutation properties in Eq. (19) instead—the underlying degrees of freedom are not fermions.

If the initial state is $|n\rangle$, i.e., the occupation number of state $|n\rangle$ is 1 while those of the other states are 0,

$$\begin{aligned} &x^{(n)}p_y^{(n)} - y^{(n)}p_x^{(n)} = 1 \\ &x^{(m)}p_y^{(m)} - y^{(m)}p_x^{(m)} = 0 \quad (m \neq n) \end{aligned} \quad , \tag{23}$$

the initial condition at time $t=0$ is then constructed as

$$\begin{aligned} &x^{(n)}(0) = \cos\theta \\ &y^{(n)}(0) = -\sin\theta \\ &p_x^{(n)}(0) = \sin\theta \\ &p_y^{(n)}(0) = \cos\theta \\ &x^{(m)}(0) = 0 \\ &y^{(m)}(0) = 0 \\ &p_x^{(m)}(0) = 0 \\ &p_y^{(m)}(0) = 0 \qquad (m \neq n) \end{aligned} \quad . \tag{24}$$

Here $\theta$ can be any real number between 0 and $2\pi$. More generally, if the normalized initial state is

$$|\Psi\rangle = \sum_{n=1}^{F} c_n |n\rangle \quad , \tag{25}$$

where $\{c_n = |c_n| e^{i\theta_n}\}$ are complex numbers that satisfy $\sum_{n=1}^{F} |c_n|^2 = 1$, one has

$$\begin{aligned} &x^{(n)}p_y^{(n)} - y^{(n)}p_x^{(n)} = |c_n|^2 \\ &x^{(n)}p_y^{(m)} - y^{(n)}p_x^{(m)} = |c_n||c_m|\cos(\theta_n - \theta_m) \quad (m \neq n) \end{aligned} \quad . \tag{26}$$

The initial condition at time $t=0$ is then chosen as

$$\begin{aligned} &x^{(n)}(0) = |c_n|\cos\theta_n \\ &y^{(n)}(0) = -|c_n|\sin\theta_n \\ &p_x^{(n)}(0) = |c_n|\sin\theta_n \\ &p_y^{(n)}(0) = |c_n|\cos\theta_n \quad (\forall n \in \{1, 2, \cdots, F\}) \end{aligned} \quad . \tag{27}$$

## 2. Mapping model II

Using the analogy with the quantum-classical correspondence for two non-commutable operators, one yields the classical Hamiltonian of Model II of Paper I[4]

$$H=\sum_{n,m=1}^{F}\frac{1}{2}\left(x^{(n)}x^{(m)}+p^{(n)}p^{(m)}\right)H_{nm} \quad . \tag{28}$$

Model II is closely related to the conventional Meyer-Miller mapping model[3, 5, 11].

If the initial state is $|n\rangle$, the occupation number representation is

$$\begin{aligned}&\frac{1}{2}\left(\left(x^{(n)}\right)^2+\left(p^{(n)}\right)^2\right)=1\\&\frac{1}{2}\left(\left(x^{(m)}\right)^2+\left(p^{(m)}\right)^2\right)=0 \quad (m\neq n)\end{aligned} \quad . \tag{29}$$

The initial condition at time $t=0$ is then constructed as

$$\begin{aligned}&x^{(n)}(0)=\sqrt{2}\cos\theta\\&p^{(n)}(0)=\sqrt{2}\sin\theta\\&x^{(m)}(0)=0\\&p^{(m)}(0)=0 \qquad (m\neq n)\end{aligned} \quad , \tag{30}$$

Here $\theta$ can be any real number between 0 and $2\pi$ . More generally, given that the initial state as Eq. (25), one obtains

$$\begin{aligned}&\frac{1}{2}\left(\left(x^{(n)}\right)^2+\left(p^{(n)}\right)^2\right)=|c_n|^2\\&\frac{1}{2}\left(x^{(n)}x^{(m)}+p^{(n)}p^{(m)}\right)=|c_n||c_m|\cos\left(\theta_n-\theta_m\right) \quad (m\neq n)\end{aligned} \quad . \tag{31}$$

The initial condition at $t=0$ is then

$$\begin{aligned}&x^{(n)}(0)=\sqrt{2}|c_n|\cos\theta_n\\&p^{(n)}(0)=\sqrt{2}|c_n|\sin\theta_n \quad \left(\forall n\in\{1,2,\cdots,F\}\right)\end{aligned} \quad . \tag{32}$$

### 3. Mapping model III

Making the analogy with the classical vector, one derives the classical Hamiltonian of Model III (in paper I[4]),

$$H=\sum_{n}\frac{1}{4}\left(\left(x^{(n)}+p_y^{(n)}\right)^2+\left(y^{(n)}-p_x^{(n)}\right)^2\right)H_{nn}+\sum_{n<m}\left(x^{(n)}p_y^{(m)}-y^{(m)}p_x^{(n)}\right)H_{nm} \quad . \tag{33}$$

When the initial state is $|n\rangle$, i.e.,

$$\begin{aligned}&\frac{1}{4}\left(\left(x^{(n)}+p_y^{(n)}\right)^2+\left(y^{(n)}-p_x^{(n)}\right)^2\right)=1\\&\frac{1}{4}\left(\left(x^{(m)}+p_y^{(m)}\right)^2+\left(y^{(m)}-p_x^{(m)}\right)^2\right)=0\quad(m\neq n)\end{aligned}\quad, \tag{34}$$

the initial condition at time $t=0$ can be constructed the same as Eq. (24). Similarly, When Eq. (25) is the initial state, its initial condition can be set the same as Eq. (27).

### 4. Mapping model IV

Similarly, the classical Hamiltonian of Model IV (in Paper I[4]) is

$$\begin{aligned}H=&\sum_{n=1}^{F}\frac{\left(x^{(n)}+p_y^{(n)}\right)^2+\left(y^{(n)}-p_x^{(n)}\right)^2}{4}H_{nn}\\&+\sum_{n<m}\left(\left(x^{(n)}x^{(m)}+p_x^{(n)}p_x^{(m)}\right)+\left(y^{(n)}y^{(m)}+p_y^{(n)}p_y^{(m)}\right)\right)H_{nm}\end{aligned}\quad. \tag{35}$$

When the initial state is $|n\rangle$, the occupation number representation is the same as Eq. (34) and the corresponding initial condition at time $t=0$ is set the same as Eq. (24). Similarly, the initial condition is given by Eq. (27) when Eq. (25) is the initial state.

Simplectic algorithms have been developed for the equations of motion of these mapping Hamiltonians in Paper I[4], which we implement for the numerical examples in Section IV.

## IV. Numerical examples

Following the isomorphism in Section II, we implement the classical mapping models for the multi-state Hamiltonian [Eq. (1)] to study the second-quantized many-electron Hamiltonian problems.

### 1. Resonant level model

Here we first apply the four classical mapping models to study population dynamics for a multi-state Hamiltonian operator, where either the pure state or the mixed state is considered as the initial density. We further show that the conventional resonant level (Landauer) model[6, 7] for a quantum dot state coupled to two electrodes is isomorphic to the same multi-state Hamiltonian. All classical mapping models for the multi-state Hamiltonian operator are then useful for studying nonequilibrium dynamics of the resonant level model.

## 1A. Multi-state Hamiltonian

Consider the multi-state Hamiltonian operator

$$\begin{aligned}\hat{H} &= \varepsilon_0 |0\rangle\langle 0| + \sum_{\substack{k=1\\ \text{Left}}}^{N} \varepsilon_k |k\rangle\langle k| + \sum_{\substack{k=1\\ \text{Left}}}^{N} t_k \left(|0\rangle\langle k| + |k\rangle\langle 0|\right) \\ &+ \sum_{\substack{m=1\\ \text{Right}}}^{N} \tilde{\varepsilon}_m |m\rangle\langle m| + \sum_{\substack{m=1\\ \text{Right}}}^{N} \tilde{t}_m \left(|0\rangle\langle m| + |m\rangle\langle 0|\right)\end{aligned} , \tag{36}$$

as depicted in Fig. 1. Here state $|0\rangle$ is coupled with two different sets of states, $\{|k\rangle\}$ on the left and $\{|m\rangle\}$ on right. All parameters in Eq. (36) are real and given by

$$\begin{aligned} \varepsilon_k &= \mu_L + \left(k - \frac{N}{2}\right)\Delta\varepsilon \quad (k = 1, \cdots, N) \\ \tilde{\varepsilon}_m &= \mu_R + \left(m - \frac{N}{2}\right)\Delta\varepsilon \quad (m = 1, \cdots, N) \\ \mu_L &= -\frac{E_V}{2}, \quad \mu_R = \frac{E_V}{2} \end{aligned} \tag{37}$$

$$\begin{aligned} t_k &= \sqrt{\frac{\Gamma\,\Delta\varepsilon}{4\pi\left(1+e^{A(\varepsilon_k - B/2)}\right)\left(1+e^{-A(\varepsilon_k + B/2)}\right)}} \quad (k = 1, \cdots, N) \\ \tilde{t}_m &= \sqrt{\frac{\Gamma\,\Delta\varepsilon}{4\pi\left(1+e^{A(\tilde{\varepsilon}_m - B/2)}\right)\left(1+e^{-A(\tilde{\varepsilon}_m + B/2)}\right)}} \quad (m = 1, \cdots, N) \end{aligned} \tag{38}$$

$$A = \frac{5}{\Gamma}, \quad B = 20\,\Gamma \tag{39}$$

$$\Delta\varepsilon = 0.0075\,\Gamma, \quad N = 400 \tag{40}$$

Set $\Gamma = 1$ in Eqs. (36)-(40). State $|0\rangle$ in Eq. (36) acts as a bridge to allow population transfer between the left set of states $\{|k\rangle\}$ and the right set of states $\{|m\rangle\}$. We set $\varepsilon_0 = -3\Gamma = -3$ for the examples in Section IV-1A. Below we focus on the flux of the population.

Given the initial density $\hat{\rho}_0$, the total population in the left set of states $\{|k\rangle\}$ at time *t* is given by

$$\sum_{\substack{k=1\\ \text{Left}}}^{N} \mathrm{Tr}\left(\hat{\rho}_0 e^{i\hat{H}t/\hbar} |k\rangle\langle k| e^{-i\hat{H}t/\hbar}\right) \quad . \tag{41}$$

It is easy to obtain the flux of the left set of states as the time derivative of Eq. (41) from Eq. (36)

$$I_L = \frac{i}{\hbar} \sum_{\substack{k=1\\ \text{Left}}}^{N} t_k \mathrm{Tr}\left[\hat{\rho}_0 e^{i\hat{H}t/\hbar} \left(|0\rangle\langle k| - |k\rangle\langle 0|\right) e^{-i\hat{H}t/\hbar}\right] \quad . \tag{42}$$

The flux of the right set of states is

$$I_R = \frac{i}{\hbar} \sum_{\substack{m=1\\ \text{Right}}}^{N} \tilde{t}_m \mathrm{Tr}\left[\hat{\rho}_0 e^{i\hat{H}t/\hbar} \left(|0\rangle\langle m| - |m\rangle\langle 0|\right) e^{-i\hat{H}t/\hbar}\right] \quad . \tag{43}$$

The total flux is then defined as

$$I_{tot} = \frac{I_R - I_L}{2} \quad . \tag{44}$$

**1) Initial density is a pure state**

The Hamiltonian of Eq. (36) is mapped onto $F = 2N + 1$ continuous degrees of freedom in the Cartesian phase space in each of the four mapping models[4]. When the initial state is a basis state in Eq. (36), the corresponding initial condition is constructed by either Eq. (24) or Eq. (30) in the four classical mapping models. The total population in the left set of states

$\{|k\rangle\}$ in Model I is represented by

$$\sum_{\substack{k=1\\ \text{Left}}}^{N} P_k = \sum_{\substack{k=1\\ \text{Left}}}^{N}\left(x^{(k)}p_y^{(k)} - y^{(k)}p_x^{(k)}\right) \quad . \tag{45}$$

The flux of the left set of states $I_L$ is the change of the total population of $\{|k\rangle\}$ over time,

$$I_L = \sum_{\substack{k=1\\ \text{Left}}}^{N} \dot{P}_k = \sum_{\substack{k=1\\ \text{Left}}}^{N}\left(\dot{x}^{(k)}p_y^{(k)} + x^{(k)}\dot{p}_y^{(k)} - \dot{y}^{(k)}p_x^{(k)} - y^{(k)}\dot{p}_x^{(k)}\right) \quad . \tag{46}$$

Similarly, the flux $I_R$ of the right set of states is

$$I_R = \sum_{\substack{m=1\\ \text{Right}}}^{N}\left(\dot{x}^{(m)}p_y^{(m)} + x^{(m)}\dot{p}_y^{(m)} - \dot{y}^{(m)}p_x^{(m)} - y^{(m)}\dot{p}_x^{(m)}\right) \quad . \tag{47}$$

The total flux $I_{tot}$ in Model I is then given by

$$\begin{aligned} I_{tot} =& \frac{1}{2}\sum_{\substack{m=1\\ \text{Right}}}^{N}\left(\dot{x}^{(m)}p_y^{(m)} + x^{(m)}\dot{p}_y^{(m)} - \dot{y}^{(m)}p_x^{(m)} - y^{(m)}\dot{p}_x^{(m)}\right) \\ & -\frac{1}{2}\sum_{\substack{k=1\\ \text{Left}}}^{N}\left(\dot{x}^{(k)}p_y^{(k)} + x^{(k)}\dot{p}_y^{(k)} - \dot{y}^{(k)}p_x^{(k)} - y^{(k)}\dot{p}_x^{(k)}\right) \end{aligned} \quad . \tag{48}$$

Substituting the Hamilton's equations of motion for the mapping Hamiltonian Eq. (22) into Eq. (48) produces the value of the total flux.

It is straightforward to obtain the expressions of the total flux in the other mapping models. While Model II produces

$$I_{tot} = \sum_{\substack{m=1\\ \text{Right}}}^{N}\left(x^{(m)}\dot{x}^{(m)} + p^{(m)}\dot{p}^{(m)}\right) - \sum_{\substack{k=1\\ \text{Left}}}^{N}\left(x^{(k)}\dot{x}^{(k)} + p^{(k)}\dot{p}^{(k)}\right) \quad , \tag{49}$$

Model III or Model IV leads to

$$\begin{aligned} I_{tot} =& \sum_{\substack{m=1\\ \text{Right}}}^{N} \frac{1}{4}\left(\left(x^{(m)} + p_y^{(m)}\right)\left(\dot{x}^{(m)} + \dot{p}_y^{(m)}\right) + \left(y^{(m)} - p_x^{(m)}\right)\left(\dot{y}^{(m)} - \dot{p}_x^{(m)}\right)\right) \\ & -\sum_{\substack{k=1\\ \text{Left}}}^{N} \frac{1}{4}\left(\left(x^{(k)} + p_y^{(k)}\right)\left(\dot{x}^{(k)} + \dot{p}_y^{(k)}\right) + \left(y^{(k)} - p_x^{(k)}\right)\left(\dot{y}^{(k)} - \dot{p}_x^{(k)}\right)\right) \end{aligned} \quad . \tag{50}$$

Choose the parameter $E_V = 4\Gamma = 4$ in Eq. (37). Consider that the initial state is the 200-

th basis state of the left set of states $\{\,|k\rangle\,\}$ or the initial density is $|k=200\rangle\langle k=200|$. The total flux as a function of time is depicted in Fig. 2a. Fig. 2b then presents the total flux when the initial state is the 200-th basis state of the right set, i.e., the initial density is $|m=200\rangle\langle m=200|$. In either case the total flux becomes a constant in long time, which indicates that a steady state appears. Fig. 2 demonstrates that all the four classical mapping models reproduce the same results as calculated in quantum mechanics. Note that the total flux given by each mapping model in either panel of Fig. 2 is efficiently obtained by using only one classical trajectory.

**2) Initial density is a mixed state**

Consider the initial density

$$\hat{\rho}_0 = \sum_{\substack{k=1 \\ \text{Left}}}^{N} f_L^{(k)} |k\rangle\langle k| + \sum_{\substack{m=1 \\ \text{Right}}}^{N} \tilde{f}_R^{(m)} |m\rangle\langle m| \tag{51}$$

with

$$f_L^{(k)} = \frac{1}{1+e^{\beta(\varepsilon_k-\mu_L)}}\ , \quad \tilde{f}_R^{(m)} = \frac{1}{1+e^{\beta(\tilde{\varepsilon}_m-\mu_R)}} \quad . \tag{52}$$

Here $\beta = 1/k_B T$ is the inverse temperature. Note that the mixed state in Eq. (51) can be decomposed into $2N$ pure states. Assume that $I_{tot}^{(k)}$ is the total flux when the initial density is $|k\rangle\langle k|$ of the left set, while $\tilde{I}_{tot}^{(m)}$ is the total flux when the initial density is $|m\rangle\langle m|$ of the right set. The total flux for the initial density Eq. (51) is then given by

$$\langle I_{tot}\rangle = \sum_{k=1}^{N} f_L^{(k)} I_{tot}^{(k)} + \sum_{m=1}^{N} \tilde{f}_R^{(m)} \tilde{I}_{tot}^{(m)} \quad . \tag{53}$$

Because each of $\{I_{tot}^{(k)}\}$ or of $\{\tilde{I}_{tot}^{(m)}\}$ is obtained with only one classical trajectory, the total flux $\langle I_{tot}\rangle$ in Eq. (53) then employs only $2N$ independent classical trajectories. The initial condition of each classical trajectory reads either Eq. (27) or Eq. (32) in the four mapping models. Evaluation of $I_{tot}^{(k)}$ or $\tilde{I}_{tot}^{(m)}$ takes the same procedure as presented in the previous case. More generally, the number of classical trajectories involved in each mapping model is the same as the number of pure states in the decomposition of the initial density.

Fig. 3 presents the total flux $\langle I_{tot} \rangle$ as a function of time for $E_V = 4\Gamma = 4$ and $\beta = 5/\Gamma = 5$. Comparison of Fig. 3 to Fig. 2 shows that the steady state value of the total flux is reached within even shorter time when the initial density is a mixed state (Eq. (51)). Eq. (53) suggests that the total flux $\langle I_{tot} \rangle$ in Fig. 3 reflects a collective behavior of the results in Fig. 2 from different initial basis states.

Fig. 4 then demonstrates how the steady state value of the total flux $\langle I_{tot} \rangle$ varies with the value of $E_V$ for $\beta = 5$ and for $\beta = 1$. In either case the total flux $\langle I_{tot} \rangle$ saturates when $E_V$ is large. The $\langle I_{tot} \rangle$ - $E_V$ curve demonstrates an "S"-shape that becomes more distinct as the temperature decreases (i.e., the value of $\beta$ increases). While the steady state value of the total flux increases as the temperature increases for small values of $E_V$, the behavior is reversed for large values of $E_V$. This interesting crossover behavior is depicted in Fig. 4c.

As shown in Figs. 3-4, each of the four classical mapping models leads to the exact results in quantum mechanics by using only $2N = 800$ classical trajectories.

### 1B. Second-quantized many-electron Hamiltonian with only 1-electron interactions

The resonant level (Landauer) model[6, 7] for a quantum dot state coupled to two electrodes is employed as a proof-of-concept example. Its Hamiltonian for quantum transport is

$$\hat{H} = \varepsilon_0 \hat{\alpha}_0^+ \hat{\alpha}_0 + \sum_{\substack{k=1 \\ \text{Left}}}^{N} \varepsilon_k \hat{\alpha}_k^+ \hat{\alpha}_k + \sum_{\substack{k=1 \\ \text{Left}}}^{N} t_k \left( \hat{\alpha}_0^+ \hat{\alpha}_k + \hat{\alpha}_k^+ \hat{\alpha}_0 \right) + \sum_{\substack{m=1 \\ \text{Right}}}^{N} \tilde{\varepsilon}_m \hat{\alpha}_m^+ \hat{\alpha}_m + \sum_{\substack{m=1 \\ \text{Right}}}^{N} \tilde{t}_m \left( \hat{\alpha}_0^+ \hat{\alpha}_m + \hat{\alpha}_m^+ \hat{\alpha}_0 \right) \quad . \tag{54}$$

where $\varepsilon_0$ is the energy of the isolated quantum dot, $\varepsilon_k$ (or $\tilde{\varepsilon}_m$) is the energy associated with the $k$-th (or $m$-th) electrode mode of the left (or right) lead, and $t_k = \sqrt{J_L(\varepsilon_k)\Delta\varepsilon/2\pi}$ (or $\tilde{t}_m = \sqrt{J_R(\varepsilon_m)\Delta\varepsilon/2\pi}$) is the coupling strength (hybridization) between the quantum dot and the $k$-th (or $m$-th) electrode mode of the left (or right) lead, which is determined from the spectral density

$$J_{L/R}(\varepsilon) = \frac{\Gamma_{L/R}}{\left(1 + e^{A(\varepsilon - B/2)}\right)\left(1 + e^{-A(\varepsilon + B/2)}\right)} \quad . \tag{55}$$

Consider zero initial dot population

$$\left\langle \hat{\alpha}_0^+(0) \hat{\alpha}_0(0) \right\rangle = 0 \quad , \tag{56}$$

no correlation at $t = 0$

$$\left\langle \hat{\alpha}_k^+(0) \hat{\alpha}_0(0) \right\rangle = \left\langle \hat{\alpha}_0^+(0) \hat{\alpha}_k(0) \right\rangle = 0 \quad , \tag{57}$$

and a Fermi-Dirac distribution for the leads' populations

$$\left\langle \hat{\alpha}_k^+(0) \hat{\alpha}_{k'}(0) \right\rangle = \delta_{kk'} \Big/ \left(1 + e^{\beta(\varepsilon_k - \mu_{L/R})}\right) \quad . \tag{58}$$

$k$ and $k'$ in Eq. (57) or (58) can be any electrode mode of the left (or right) lead. The left or right current as a function of time is given by

$$\left\langle \tilde{I}_{L/R}(t) \right\rangle = -e \frac{d}{dt} \left\langle \sum_{k \in L/R} \hat{\alpha}_k^+ \hat{\alpha}_k \right\rangle \quad , \tag{59}$$

where $e$ is the elementary charge. (Set $e = 1$ in the paper.) The total current is defined as half the difference between the left current and the right one, i.e.,

$$\left\langle \tilde{I}_{tot}(t) \right\rangle = \frac{\left\langle \tilde{I}_L(t) \right\rangle - \left\langle \tilde{I}_R(t) \right\rangle}{2} \tag{60}$$

It is straightforward to follow Section II to show that the second-quantized many-electron Hamiltonian [Eq. (54)] with the initial condition [Eqs. (56)-(58)] is isomorphic to the multi-state Hamiltonian [Eq. (36)] with the initial density [Eqs. (51)-(52)]. One can further demonstrate that the relation between the total current of Eq. (60) and the total flux of Eq. (53) is

$$\left\langle \tilde{I}_{tot}(t)\right\rangle = e\left\langle I_{tot}\right\rangle \quad . \tag{61}$$

Models I-IV may then be applied to the resonance level model in the same procedure as that described in Section IV-1A. The parameters of Eq. (40) are sufficient to obtain converged results up to $t = 20\hbar/\Gamma$ for which a steady state appears.

Fig. 5 shows that each of four classical mapping models is able to efficiently reproduce the exact results derived in the literature[6, 7] by using only $2N = 800$ trajectories. For comparison, $10^5 \sim 10^6$ trajectories were employed in Swenson *et al.*'s semiclassical mapping model[7] or in Li *et al.*'s quasiclassical approach[12] for the resonance level model. Swenson *et al.*'s semiclassical mapping model[7] is not exact, as suggested by comparison of its results to exact ones[7]. Li *et al.*'s quasiclassical approach[12] employed a semiclassical second-quantized many-electron Hamiltonian[8] that is identical to the mapping Hamiltonian of Model I. Fig. 5 suggests that the classical mapping models with corresponding initial conditions are much more efficient for treating the second-quantized many-electron Hamiltonian (at least when only 1-electron interactions are involved).

Interestingly, the isomorphism between the multi-state Hamiltonian [Eq. (36)] and the second-quantized many-electron Hamiltonian [Eq. (54)] indicates that the total current of the resonance level model can demonstrate an "S"-shape voltage dependence that becomes more distinct as the temperature decreases (or $\beta$ increases). The relation between the steady state value of the total current and the temperature (for the resonance level model) also has the same crossover behavior as depicted in Fig. 4.

## 2. Two electrons and four spin-orbitals

Use the model system—2 electrons and 4 spin-orbitals (2 spatial orbitals, each with $\bar{\alpha}$ and $\bar{\beta}$ spin) of Ref. [8] as a proof-of-concept example for the second-quantized many-electron Hamiltonian with 2-electron interactions. The parameters of Eq. (8) are given by Eq. (3.3) or Eq. (3.4) of Ref. [8]. The many-electron basis (i.e., 2-electron basis in this case) is $\left\{ |j=1\rangle \equiv |1\bar{\alpha}1\bar{\beta}\rangle,\ |j=2\rangle \equiv |1\bar{\alpha}2\bar{\beta}\rangle,\ |j=3\rangle \equiv |2\bar{\alpha}1\bar{\beta}\rangle,\ |j=4\rangle \equiv |2\bar{\alpha}2\bar{\beta}\rangle \right\}$. The second-quantized many-electron Hamiltonian Eq. (8) is then represented by

$$\begin{array}{cc} & \begin{array}{cccc} |1\rangle & |2\rangle & |3\rangle & |4\rangle \end{array} \\ \hat{H} = & \begin{pmatrix} 0.06 & 0.22 & -0.22 & 0.08 \\ 0.22 & 1.12 & -0.08 & 0.25 \\ -0.22 & -0.08 & 1.12 & -0.25 \\ 0.08 & 0.25 & -0.25 & 2.04 \end{pmatrix} \begin{array}{c} \langle 1| \\ \langle 2| \\ \langle 3| \\ \langle 4| \end{array} \end{array} , \tag{62}$$

which is a multi-state Hamiltonian operator (i.e., Eq. (1)). The classical mapping models of Section III may then be applied to Eq. (62) that is an equivalent expression of the second-quantized many-electron Hamiltonian with 2-electron interactions for the model system.

Set the initial state as a superposition of the 4 basis states

$$|\Psi\rangle = \frac{1}{\sqrt{30}}\left(|1\bar{\alpha}1\bar{\beta}\rangle + 2|1\bar{\alpha}2\bar{\beta}\rangle + 4|2\bar{\alpha}1\bar{\beta}\rangle + 3|2\bar{\alpha}2\bar{\beta}\rangle\right) . \tag{63}$$

Note that the population of state $|j\rangle$ as a function of time is a kind of correlation function

$$P_j(t) = \mathrm{Tr}\left(\hat{\rho}_0 e^{i\hat{H}t/\hbar} |j\rangle\langle j| e^{-i\hat{H}t/\hbar}\right) = \left|\langle j| e^{-i\hat{H}t/\hbar} |\Psi\rangle\right|^2 . \tag{64}$$

Its Fourier transform depicts excitations between different eigenstates. It is easy to show that the spectrum obtained from the correlation function in Eq. (64) is not always positive-definite. The Wiener-Khintchine theorem[13] does not apply. This is different from the spectrum obtained from the thermal correlation function[13]. Since the population $n_j(t)$ never decays in this case, the spectrum is then obtained from

$$\begin{aligned} I_j(\omega) &\approx \frac{1}{2\pi}\int_{-\infty}^{\infty} dt \left(P_j(t) - \langle P_j(t)\rangle\right) e^{-i\omega t} \exp\left[-\frac{t^2}{2\sigma^2}\right] \\ &= \frac{1}{\pi}\int_{0}^{\infty} dt \left(P_j(t) - \langle P_j(t)\rangle\right) e^{-i\omega t} \exp\left[-\frac{t^2}{2\sigma^2}\right] \end{aligned} , \quad (65)$$

where the Gaussian width parameter $\sigma$ is set to be large enough to accurately capture the peak positions of the spectrum of the system. When the width parameter $\sigma$ approaches infinity, Eq. (65) leads to a set of delta functions in this case. The average population $\langle P_j(t)\rangle$ in Eq. (65) is defined as

$$\langle P_j(t)\rangle = \lim_{t_{\max}\to\infty} \frac{1}{t_{\max}} \int_0^{t_{\max}} dt\, P_j(t) \quad . \quad (66)$$

As $t_{\max}$ increases, the value of $\langle P_j(t)\rangle$ becomes converged.

Choose any one of the four classical mapping models to calculate the population of $|j=2\rangle \equiv |1\bar{\alpha}2\bar{\beta}\rangle$, i.e.,

$$n_2(t) = x^{(2)}(t)\, p_y^{(2)}(t) - y^{(2)}(t)\, p_x^{(2)}(t) \quad (67)$$

for Model I,

$$n_2(t) = \frac{1}{2}\left(\left(x^{(2)}(t)\right)^2 + \left(p^{(2)}(t)\right)^2\right) \quad (68)$$

for Model II, or

$$n_2(t) = \frac{1}{4}\left(\left(x^{(2)}(t) + p_y^{(2)}(t)\right)^2 + \left(y^{(2)}(t) - p_x^{(2)}(t)\right)^2\right) \quad (69)$$

for Model III or IV. Only one trajectory is propagated for evaluating $n_2(t)$. The Fourier transform of the population of $\left|1\bar{\alpha}2\bar{\beta}\right\rangle$ leads to the spectrum as shown in Fig. 6. (The converged result is obtained with $\sigma = 600$ and $t_{\max} = 1600$.) The spectrum in the negative frequency regime is not demonstrated because it is simply symmetric to that in the positive frequency domain. The frequencies of Fig. 6 read $\{0.9441, 1.0506, 1.0594, 1.1450, 1.1538, 2.2045\}$. Note that the four energy eigenvalues of the Hamiltonian Eq. (62) are $\{-0.0194, 1.040, 1.1344, 2.185\}$. The total number of excitations between any two eigenstates is $4\times3=12$. The exact frequencies are { $\pm0.9441$, ±1.0506, ±1.0594, ±1.1450, ±1.1538, $\pm2.2045$ }, the positive values of which are well reproduced in Fig. 6. (See Appendix for more discussion on obtaining exact energy eigenvalues of the system.)

## V. Conclusions

We introduce an isomorphism between the multi-state Hamiltonian and the second-quantized many-electron Hamiltonian (with 1-electron interactions). This sets the scene such that all methods developed for the former can be applied to the latter, and vice versa. When only 1-electron interactions are employed, the anticommutation relations [Eq. (17)] are not those of fermionic operators. The resonance model for quantum transport is used as a proof-of-concept example. The series of classical mapping models proposed in the unified theoretical framework of Paper I[4] are demonstrated to be efficient for studying the steady state current.

The isomorphism no longer exists when 2-electron interactions are involved in the second-quantized many-electron Hamiltonian. Many-electron states can be employed to construct the multi-state Hamiltonian operator as an equivalent expression of the second-quantized many-electron Hamiltonian for the system. A model system consisted of 2 electrons and 4 spin-orbitals in Ref. [8] is taken as another proof-of-concept example, where the classical mapping models of Paper I[4] reproduce exact excitation frequencies or exact energy eigenvalues with only one trajectory. This implies that all methods for the multi-state Hamiltonian are useful to study nonequilibrium quantum dynamics models in condensed phase (such as the Anderson impurity model[12, 14]) and the electronic structure problems (e.g., real time time-dependent Hartree-Fock, real time time-dependent density functional theory, *etc*. [15]) for molecular systems. It will be interesting to investigate along these directions in our future work.

Finally, we note that the strategy introduced in Sections I-II (and in Paper I[4]) are useful for studying the underlying degrees of freedom for general quasi-particles. Define the creation and annihilation operators for the quasi-particles as in Eq. (4) such that the space is complete for all (combined) excitations. Commutation and anti-commutation relations are then naturally constructed, from which (dynamics) methods may be rationally constructed for the quasi-particles.

**Acknowledgement**

J. L. thanks Wenjian Liu for several useful discussions on the second quantization. This work was supported by the National Natural Science Foundation of China (NSFC) Grants No. 21373018 and No. 21573007, by the Recruitment Program of Global Experts, by Specialized Research Fund for the Doctoral Program of Higher Education No. 20130001110009, by the Ministry of Science and Technology of China (MOST) Grant No. 2016YFC0202803, and by Special Program for Applied Research on Super Computation of the NSFC-Guangdong Joint Fund (the second phase). We acknowledge the Beijing and Tianjin supercomputer centers for providing computational resources.

## Appendix: Spectrum for energy eigenvalues from the classical mapping models

The classical mapping models also offer an approach to obtain energy eigenvalues for the multi-state Hamiltonian or for the second-quantized many-electron Hamiltonian. Recall the well-known relation to obtain energy eigenvalues

$$\begin{aligned}\langle\phi|\delta(E-\hat{H})|\Psi\rangle &= \frac{1}{2\pi\hbar}\int_{-\infty}^{\infty} dt\, e^{iEt/\hbar}\langle\phi|e^{-i\hat{H}t/\hbar}|\Psi\rangle \\ &= \frac{1}{\pi\hbar}\int_{0}^{\infty} dt\, \mathrm{Re}\left[e^{iEt/\hbar}\langle\phi|e^{-i\hat{H}t/\hbar}|\Psi\rangle\right]\end{aligned} \quad . \tag{70}$$

It is easy to prove Eq. (70) by inserting a complete set of eigenstates of $\hat{H}$.

Still use the model system consisted of 2 electrons and 4 spin-orbitals in Ref. [8] as an example. When the initial condition is given by Eq. (63), the amplitude of $|j=2\rangle$ at time $t$ is $\langle 2|e^{-i\hat{H}t/\hbar}|\Psi\rangle$. It is expressed as

$$\langle 2|e^{-i\hat{H}t/\hbar}|\Psi\rangle = \frac{x^{(2)}(t)+p_y^{(2)}(t)}{2} + i\,\frac{p_x^{(2)}(t)-y^{(2)}(t)}{2} \tag{71}$$

in Model I, III, or IV and

$$\langle 2|e^{-i\hat{H}t/\hbar}|\Psi\rangle = \frac{1}{\sqrt{2}}\left(x^{(2)}(t)+i\, p_x^{(2)}(t)\right) \tag{72}$$

in Model II. Since the amplitude $\langle 2|e^{-i\hat{H}t/\hbar}|\Psi\rangle$ never decays in this case, the Gaussian-smoothing technique used in Eq. (65) is also applied to Eq. (70) for obtaining the spectrum for energy eigenvalues. Only one trajectory is propagated for evaluating $\langle 2|e^{-i\hat{H}t/\hbar}|\Psi\rangle$ in any one of Models I-IV. Use Model I for demonstration, while Models II-IV lead to the same results. The Gaussian parameter is $\sigma = 600$ and the total length of propagation time is 1600, which are enough for obtaining converged results for the peak positions. The corresponding spectrum of $\langle 2|e^{-i\hat{H}t/\hbar}|\Psi\rangle$ is depicted in Fig. 7, in which the peak positions well reproduce the exact energy eigenvalues of the system $\{-0.0194, 1.040, 1.1344, 2.185\}$.

**Figure Captions**

**Fig. 1** (Color). Schematic representation of the multi-state system of Eq. (36). State $|0\rangle$ is coupled with the left set of states $\{\,|k\rangle\,\}$ and the right set of states $\{\,|m\rangle\,\}$.

**Fig. 2** (Color). The total flux $I$ as a function of time for $E_V = 4\Gamma = 4$ and $\varepsilon_0 = -3\Gamma = -3$ for the multi-state Hamiltonian system of Eq. (36). (a) Initial state is the 200$^{th}$ basis state $|k = 200\rangle$ of the left set of states. (b) Initial state is the 200$^{th}$ basis state $|m = 200\rangle$ of the right set of states. Solid line: exact results. Solid squares: results of Model I. Solid triangles: results of Model II. Solid rhombuses: results of Model III. Solid circles: results of Model IV. ($\Gamma = 1, \hbar = 1$)

**Fig. 3** (Color). As in Fig. 2. The total flux $\langle I_{tot} \rangle$ as a function of time for $E_V = 4\Gamma = 4, \varepsilon_0 = -3\Gamma = -3$ and $\beta = 5/\Gamma = 5$ for the Hamiltonian Eq. (36) and the initial density Eq. (51).

**Fig. 4** (Color). The steady state value of the total flux $\langle I_{tot} \rangle$ as a function of $E_V$ for $\beta = 5$ (Panel a) and that for $\beta = 1$ (Panel b). Solid line: exact results. Hollow squares: results of Model I. Hollow triangles: results of Model II. Hollow rhombuses: results of Model III. Solid circles: results of Model IV. ($\Gamma = 1, \hbar = 1, \varepsilon_0 = -3\Gamma = -3$)

Panel c: Comparison of the results of Model III for $\beta = 5$ to those for $\beta = 1$. Solid line with solid circles: $\beta = 5$. Dashed line with solid squares: $\beta = 1$.

**Fig. 5** (Color). Transient currents for the resonant level model [Eq. (54)]. (a) Results for source-drain voltages $E_V$ while other parameters are $\mu_L = -\mu_R = E_V/2, \beta = 3/\Gamma$, and $\varepsilon_0 = 0$. (b) Results for different gate voltages $\varepsilon_0 = \mathrm{eV}_g$ while other parameters are $\mu_L = -\mu_R = \Gamma$ and $\beta = 3/\Gamma$. (c) Results for different temperatures while other parameters are $\mu_L = -\mu_R = \Gamma$ and $\varepsilon_0 = 0$. Solid line: exact results. Squares: Model I. Triangles: Model II. Rhombuses: Model III. Circles: Model IV. ($\Gamma = 1, \hbar = 1, e = 1$)

**Fig. 6** (Color). Spectrum for excitations obtained from the population of $|1\bar{\alpha}2\bar{\beta}\rangle$ (for the system consisted of 2 electrons and 4 spin-orbitals). The spectrum in the negative frequency regime (not shown) is symmetric to that in the positive frequency domain. Model I is used for demonstration, while other classical mapping models lead to the same results.

**Fig. 7** (Color). Spectrum for energy eigenvalues obtained from the amplitude $\langle 2|e^{-iHt/\hbar}|1\bar{\alpha}2\bar{\beta}\rangle$ (for the system consisted of 2 electrons and 4 spin-orbitals). Model I is implemented, while other classical mapping models produce the same results.

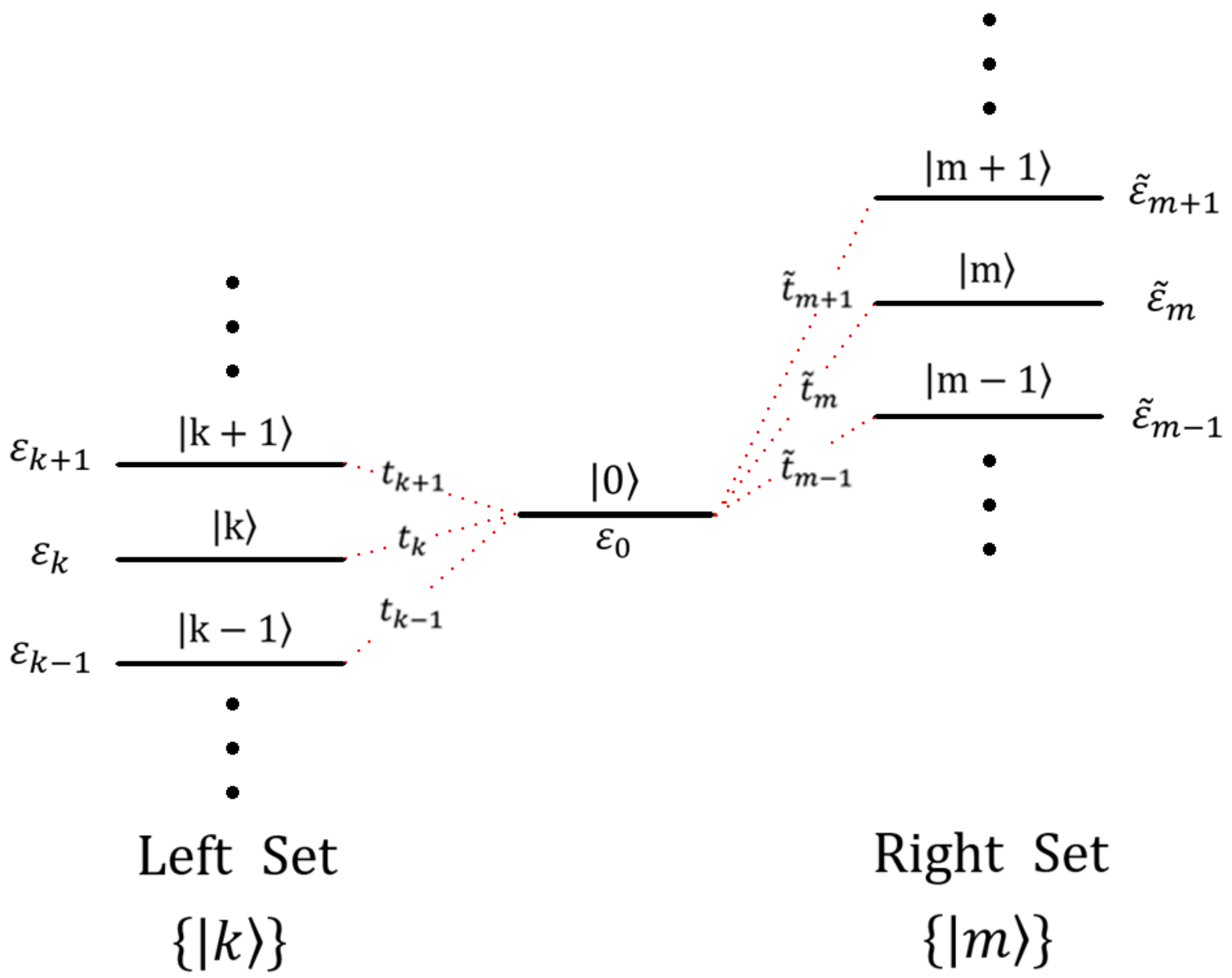

**Fig. 1**

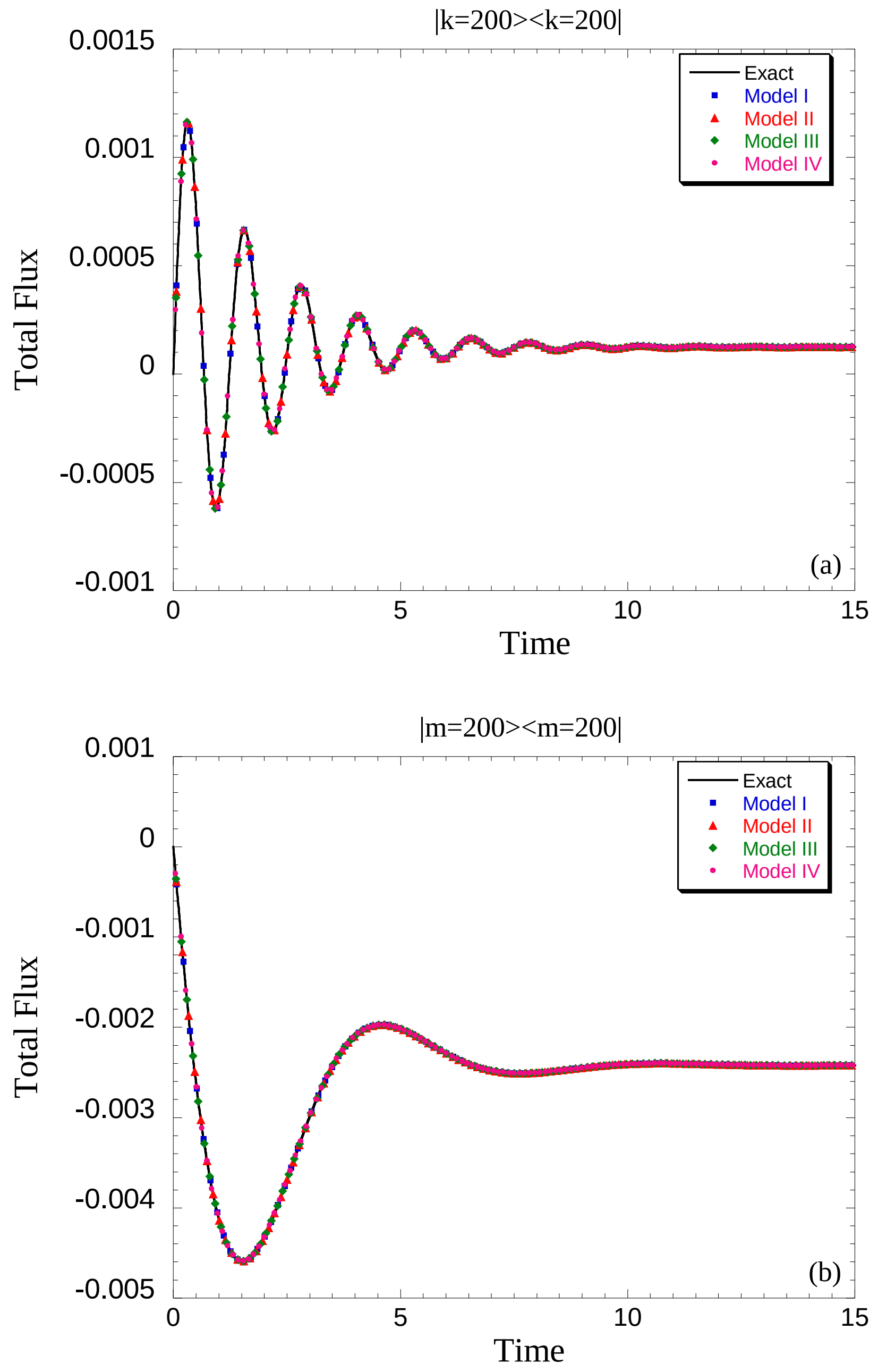
|k=200><k=200|
Total Flux
0.0015
0.001
0.0005
0
-0.0005
-0.001
Time
0
5
10
15
Exact
Model I
Model II
Model III
Model IV
(a)
|m=200><m=200|
Total Flux
0.001
0
-0.001
-0.002
-0.003
-0.004
-0.005
Time
0
5
10
15
Exact
Model I
Model II
Model III
Model IV
(b)

**Fig. 2**

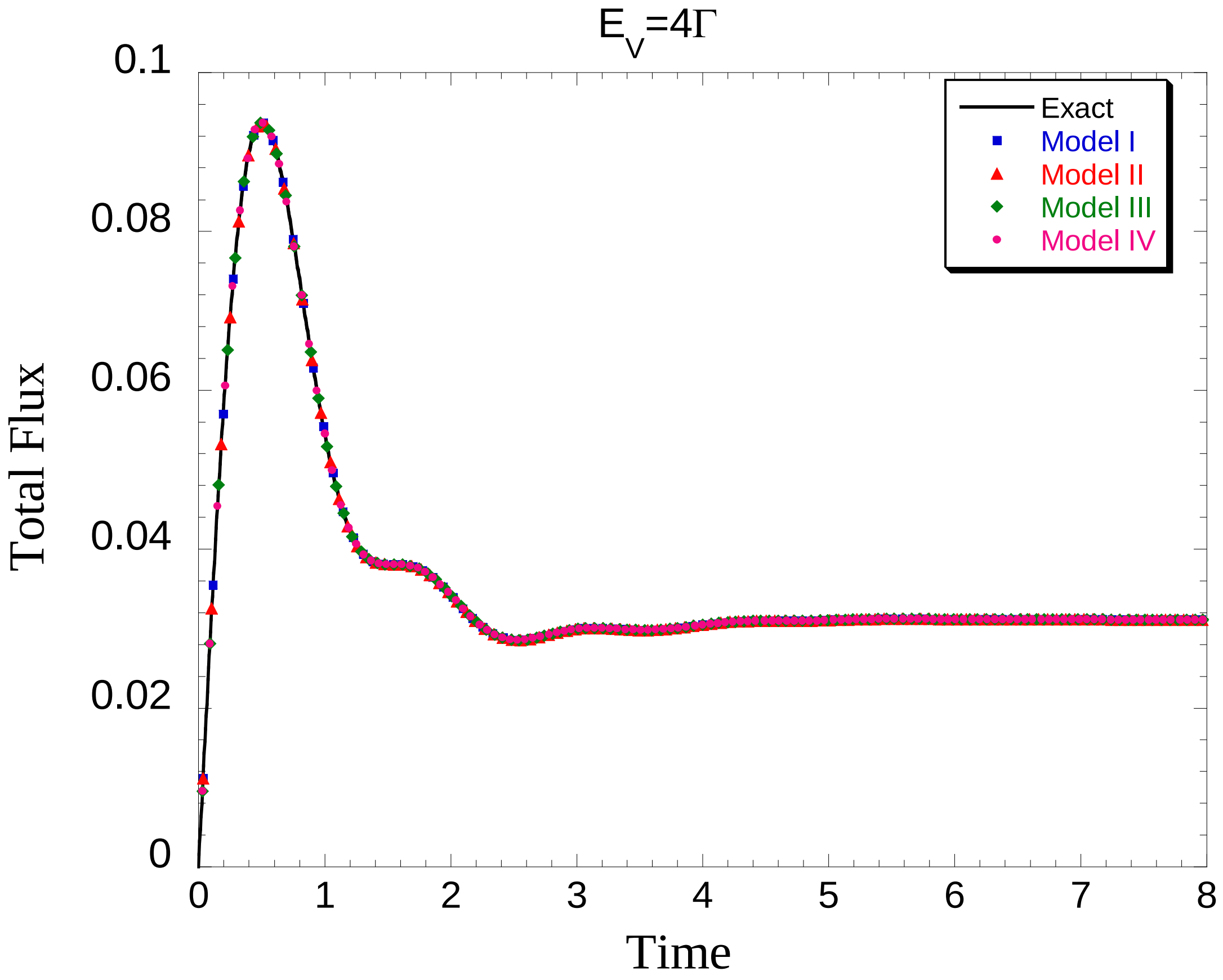
$E_V=4\Gamma$
Total Flux
Time
Exact
Model I
Model II
Model III
Model IV
0
0.02
0.04
0.06
0.08
0.1
0 1 2 3 4 5 6 7 8

**Fig. 3**

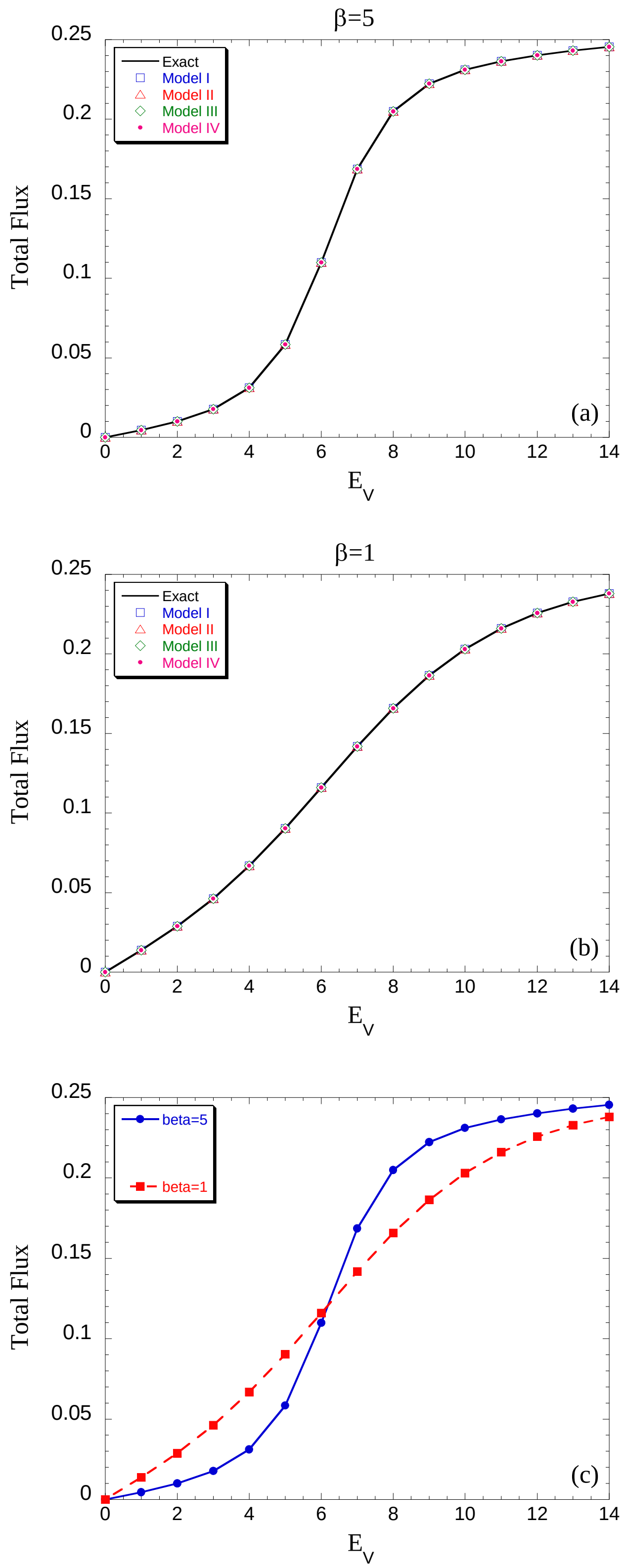

**Fig. 4**

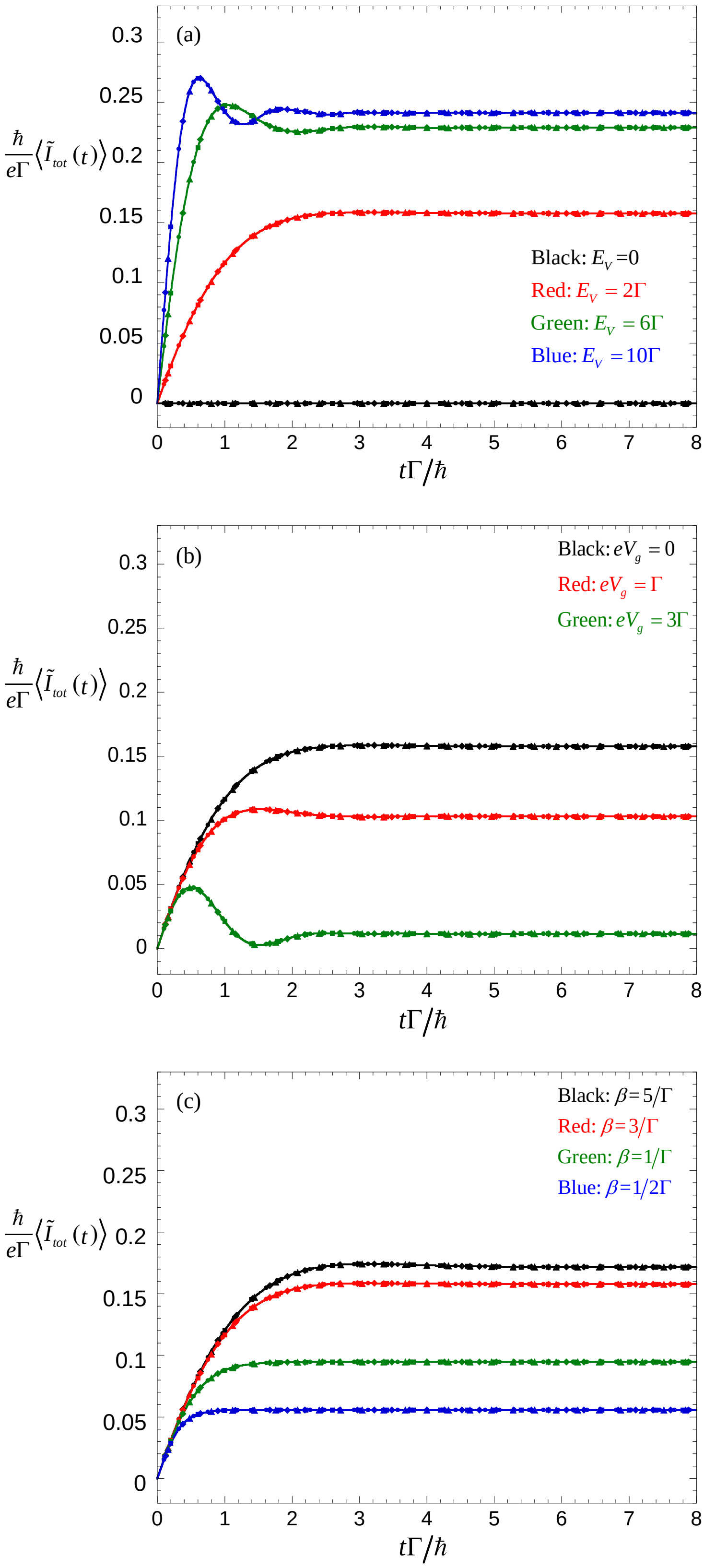

(a)
0.3
0.25
0.2
0.15
0.1
0.05
0
$\frac{\hbar}{e\Gamma}\langle \tilde{I}_{tot}(t)\rangle$
Black: $E_V = 0$
Red: $E_V = 2\Gamma$
Green: $E_V = 6\Gamma$
Blue: $E_V = 10\Gamma$
0 1 2 3 4 5 6 7 8
$t\Gamma/\hbar$
(b)
Black: $eV_g = 0$
Red: $eV_g = \Gamma$
Green: $eV_g = 3\Gamma$
$t\Gamma/\hbar$
(c)
Black: $\beta = 5/\Gamma$
Red: $\beta = 3/\Gamma$
Green: $\beta = 1/\Gamma$
Blue: $\beta = 1/2\Gamma$
$t\Gamma/\hbar$

**Fig. 5**

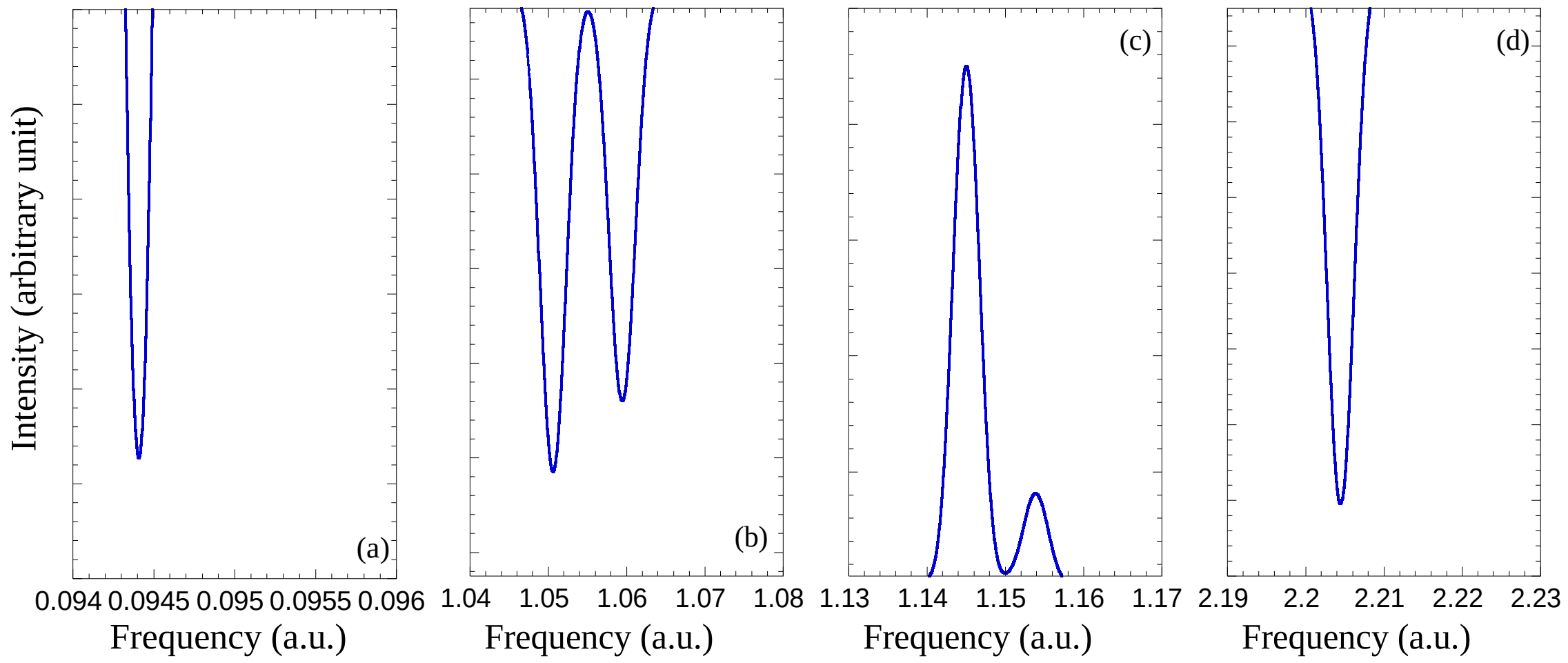
Intensity (arbitrary unit)
(a)
0.094 0.0945 0.095 0.0955 0.096
Frequency (a.u.)
(b)
1.04 1.05 1.06 1.07 1.08
Frequency (a.u.)
(c)
1.13 1.14 1.15 1.16 1.17
Frequency (a.u.)
(d)
2.19 2.2 2.21 2.22 2.23
Frequency (a.u.)

**Fig. 6**

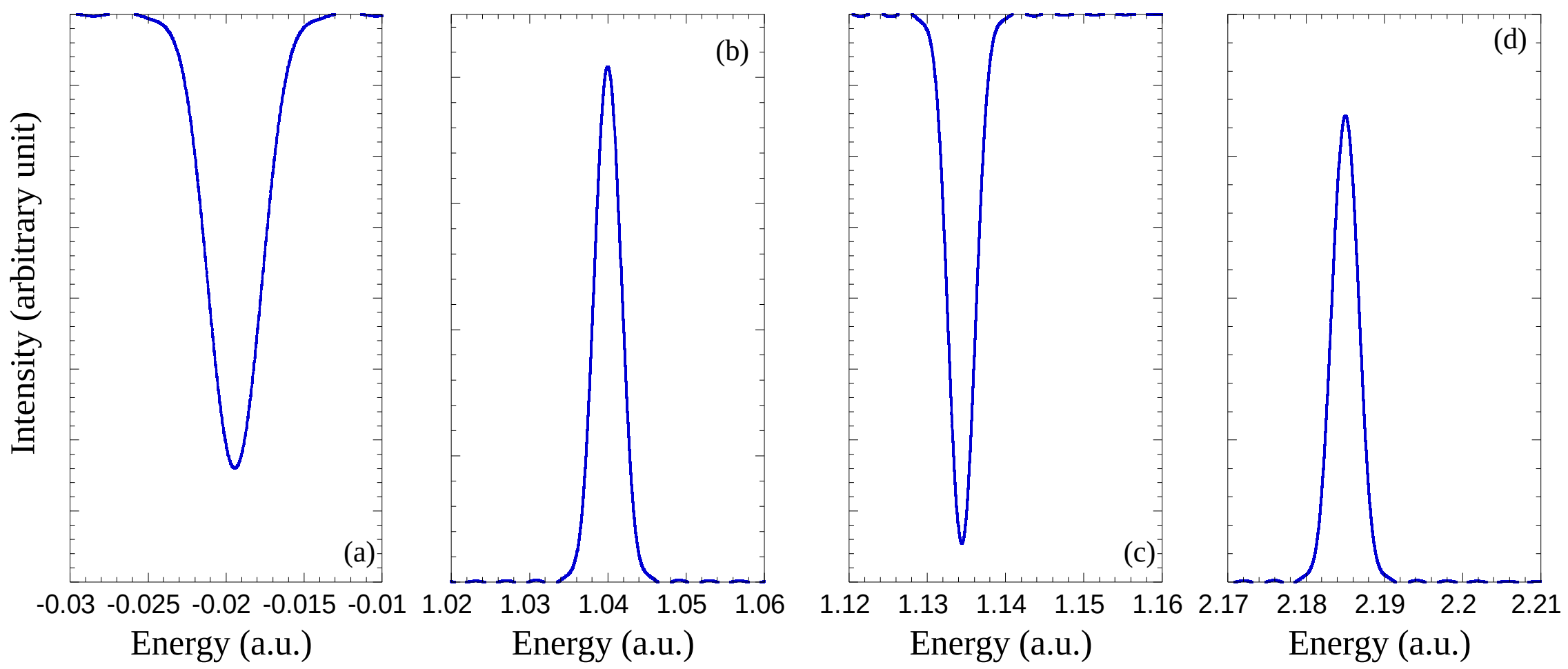

**Fig. 7**